\documentclass[a4paper,fleqn,usenatbib]{mnras}

\usepackage{newtxtext,newtxmath}

\usepackage[T1]{fontenc}
\usepackage{ae,aecompl}

\usepackage{graphicx}	
\usepackage{amsmath}	
\usepackage{amssymb}	
\usepackage{pdflscape}

\title[The Magnetic Field in the central parsec. ] {The Magnetic Field in the central parsec of the Galaxy }
\author[P. F. Roche, et al ]{  P. F. Roche$^{1}$\thanks{E-mail: p.roche1@physics.ox.ac.uk (PFR)}, E. Lopez-Rodriguez $^{2,3}$,  C.M. Telesco$^{4}$, 
R. Sch{\"o}del $^{5}$, C. Packham$^{6,7}$  \\
$^{1}$ {Astrophysics, Department of Physics, University of Oxford, DWB, Keble Road, Oxford OX1 3RH} \\
$^{2}$ {SOFIA Science Center,  NASA AmesResearch Center, Moffett Field, CA 94035, USA} \\
$^{3}$ {Dept of Astronomy, University of Texas at Austin, 1 University Station, C1400, Austin, TX 78712, USA  } \\
$^{4}$ {Department of Astronomy, University of Florida, Gainesville FL 32611, USA} \\
$^{5}$ {Instituto de Astrofisica de Andalucia (CSIC), Glorieta de la Astronomia S/N, 18008 Granada, Spain}\\
$^{6}$ {Dept. of Physics \& Astronomy, University of Texas at San Antonio,1 UTSA Circle, San Antonio, Texas, 78249, USA } \\
$^{7}$ {National Astronomical Observatory of Japan, Mitaka, Tokyo 181-8588, Japan} 
}
\date{Accepted 2018 January 10. Received 2018 January 9; in original form 2017 August 3}

\pubyear{2018}

\begin{document}
\label{firstpage}
\pagerange{\pageref{firstpage}--\pageref{lastpage}}
\maketitle


\begin{abstract}
We present a polarisation map of the warm dust emission from the minispiral  in the central parsec of the Galactic centre.  The observations were made at a wavelength of 12.5~$\mu$m with CanariCam mounted on the 10.4-m Gran Telescopio Canarias.  The magnetic field traced by the polarised emission from aligned dust grains is consistent with previous observations, but the increased resolution of the present data reveals considerably more information on the detailed structure of the B field and its correspondence with the filamentary emission seen in both mid-infrared  continuum emission and free-free emission at cm wavelengths.  The magnetic field appears to be compressed and pushed by the outflows from luminous stars in the Northern Arm, but it is not disordered by them.   We identify some magnetically coherent filaments that cross the Northern Arm at a Position Angle of $\sim 45^{\rm o}$,  and which may trace orbits inclined to the primary orientation of the Northern Arm and circumnuclear disk.  In the East-West bar, the magnetic fields implied by the polarization in the lower intensity regions lie predominantly along the bar at a Position Angle of $130 - 140^{\rm o}$.  In contrast to the Northern Arm, the brighter regions of the bar tend to have lower degrees of polarization with a greater divergence in position angle compared to the local diffuse emission.  It appears that the diffuse emission in the East-West bar traces the underlying field and that the bright compact sources are unrelated objects presumably projected onto the bar and with different field orientations. 
\end{abstract}

\section{Introduction}

The central few parsecs of the Milky Way Galaxy host a rich and complex environment, which includes the central cluster of old stars, the 4 x 10$^6$ M$_\odot$ black hole, numerous hot mass-losing stars,and a young stellar cluster together with molecular, ionised and neutral gas mixed with dust (e.g. \citealt{Genzel10}).   Observations of continuum emission at mid-infrared wavelengths trace out emission from warm dust heated by radiation from these numerous sources sources \citep{Smith90, Telesco96, Viehmann06, Lau13}.  The most prominent structure associated with mid-IR emission from warm dust coincides  with the so-called  mini-spiral, a quasi-coheremt structure seen in  radio continuum (free-free emission)  (e.g. \citealt{Gezari91}) and in ionised gas through the 12.8~$\mu$m [Ne II] fine structure line \citep{Lacy91} and in high-n H recombination lines \citep{Roberts93, Zhao10}. Radial velocity measurements of the mini-spiral have been used by a number of authors to investigate the kinematics of the ionized gas in order to study the relationship between this material and other components of the Galactic Centre.  The Western Arc of the minispiral appears to trace the inner edge of the 1.5 - 5~pc circumnuclear disk (CND) of molecular material around the Galactic Centre \citep{Serabyn85}.  The Northern Arm and at least parts of the East-West Bar may coincide with ionization fronts at the interface between an infalling neutral cloud and the ionized gas in the central cavity \citep{Telesco96}. 

The ionized gas kinematics have been fitted by both circular (\citealt{Lacy91, Serabyn88}) and elliptical Keplerian orbits (\citealt{Paumard04, Zhao09}) around the centre, possibly with an infalling component. \citealt{Irons12} argue that the Northern Arm of the minispiral connects with the Western Arc to trace an Archimedean spiral structure that may be induced by a density wave in the medium.  They  point out the similarity of the orbital inclinations of the ionized gas in the minispiral to that of the molecular material in the CND.  Regardless of the detailed orbital shapes, the minispiral appears to trace the ionized edge of a much larger structure, while the minispiral orbital plane appears to be very similar to that of the circumnuclear disk, suggesting a close relationship and perhaps a common origin. Neutral gas, traced by [O I] emission, fills in the region between the Northern Arm of the minispiral  and the North-Eastern arc of the CND \citep{Jackson93}, suggesting a direct relationship between the ionized, neutral and molecular components    in that region.  Recent observations with ALMA have shown a complex distribution of clumps emitting a range of molecular lines. While most of the molecular gas emission traces the CND, a number of clumps are detected in the region west of the Northern Arm, north of the East-West bar and extending north and west of SgrA*. The total mass of molecular gas in the central parsec appears to be a small fraction of that in the ionized gas (\citealt{Moser17}). 

Measurements of the emissive polarisation from warm, aligned dust grains at mid-IR wavelengths have been employed to estimate the direction of the component of the magnetic field in the plane of the sky and place some constraints on the likely field strength (\citealt{Aitken91, Aitken98}; \citealt{Glasse03}).  Spectropolarimetry has allowed the contribution from the absorptive interstellar polarisation to be separated from the emissive polarisation arising from   warm aligned grains in the dusty structures in the minispiral \citep{Aitken86}.   The polarization in the Northern Arm is quite ordered, suggesting that the magnetic field is strong, with a field strength of $\sim$2~mGauss, and that it lies primarily along the arc of the Northern Arm.  \citet{Aitken98} have argued that the polarization may be saturated in the Northern Arm, such that reductions in percentage polarization are related to the inclination of the field with respect to the plane on the sky; in this way, with the radial velocity measurements of \citet{Roberts93} they were able to use the observed polarization to construct a 3-D estimate of the kinematics.  The polarization in the East-West bar of the minispiral was found to be much less ordered and the relationship between the polarization and radial velocity structures is less clear.    The resolution of these observations was  diffraction-limited to 0.8 arcsec by the primary mirrors of the 4-m Anglo-Australian and UK Infrared telescopes.  The data presented here, obtained with the 10.4-m  GTC, provides an improvement in resolution of a factor of 2.

Here we present 12.5~$\mu$m  imaging polarimetry  of the central  0.75 parsec (at an assumed distance of 8~kpc; \citealt{Boehle16})  of the Galactic Centre  to investigate the polarisation and magnetic structures  at higher resolution than previous observations, and in particular to investigate the relationship between the outflow sources and the magnetic field.  

\section{Observations}

The observations were obtained with the CanariCam mid-IR multi-mode camera  (\citealt{Telesco03}) mounted on the Nasmyth A platform of the 10.4-m Gran Telescopio Canarias  between 2015 July 28 and August 02 in programme GTC6-14AGCAN.  CanariCam was used in its dual-beam polarimetric mode  (\citealt{Packham2005}) wherein  a Wollaston prism separates the o- and e-rays and the polarisation is modulated by a cryogenic CdSe half-wave plate installed in the upstream optical path. The Wollaston prism is used in conjunction with a focal plane mask consisting of three slots, each 20 arcsec long x 2 arcsec wide, and separated by 5 arcsec which prevents overlap of the images in the e- and o-rays on the 320 x 240 pixel Raytheon Si:As detector.  

The telescope position was offset  in steps of 1 arcsec between exposures, with six steps providing full sampling of the field.  The Si-6  12.5~$\mu$m filter,  was chosen as it offers good sensitivity and high polarization efficiency  while maximising the emissive polarisation component and minimising the absorptive interstellar polarization (see section 4.3).  The filter bandpass (50\% of peak transmission) is 12.08 -- 12.76~$\mu$m.  Standard  chop-nod sequences were used, with the waveplate rotated in sequence  by 0, 22.5, 45 and 67.5 degrees.  A chop-throw of 25$^{\prime\prime}$ at position angle 45 degrees east of north was used to minimise contamination from emission in the reference beam.  The pixel size of CanariCam is 0.080$^{\prime\prime}$.   Each observing block (OB) consisted of a set of six exposures shifted successively by 1 arcsec in Right Ascension, each with an on-source integration time of  126 seconds.  As the slots are 2 arcsec wide, each pixel on the sky is exposed for 256 sec in each OB.  The polarization map presented here is the accumulated signal from three observing blocks with a total  on-source  exposure time  per pixel  of 12.6 minutes.  The Galactic Centre only rises above 2 airmasses for short periods on La Palma, and although the conditions for the observations were good, the low elevation at which they were conducted resulted in the image quality of the final image being $\sim$0.45 arcsec compared to the diffraction limit of 0.3 arcsec.

\begin{figure*}
\centering
	\includegraphics[width=15cm]{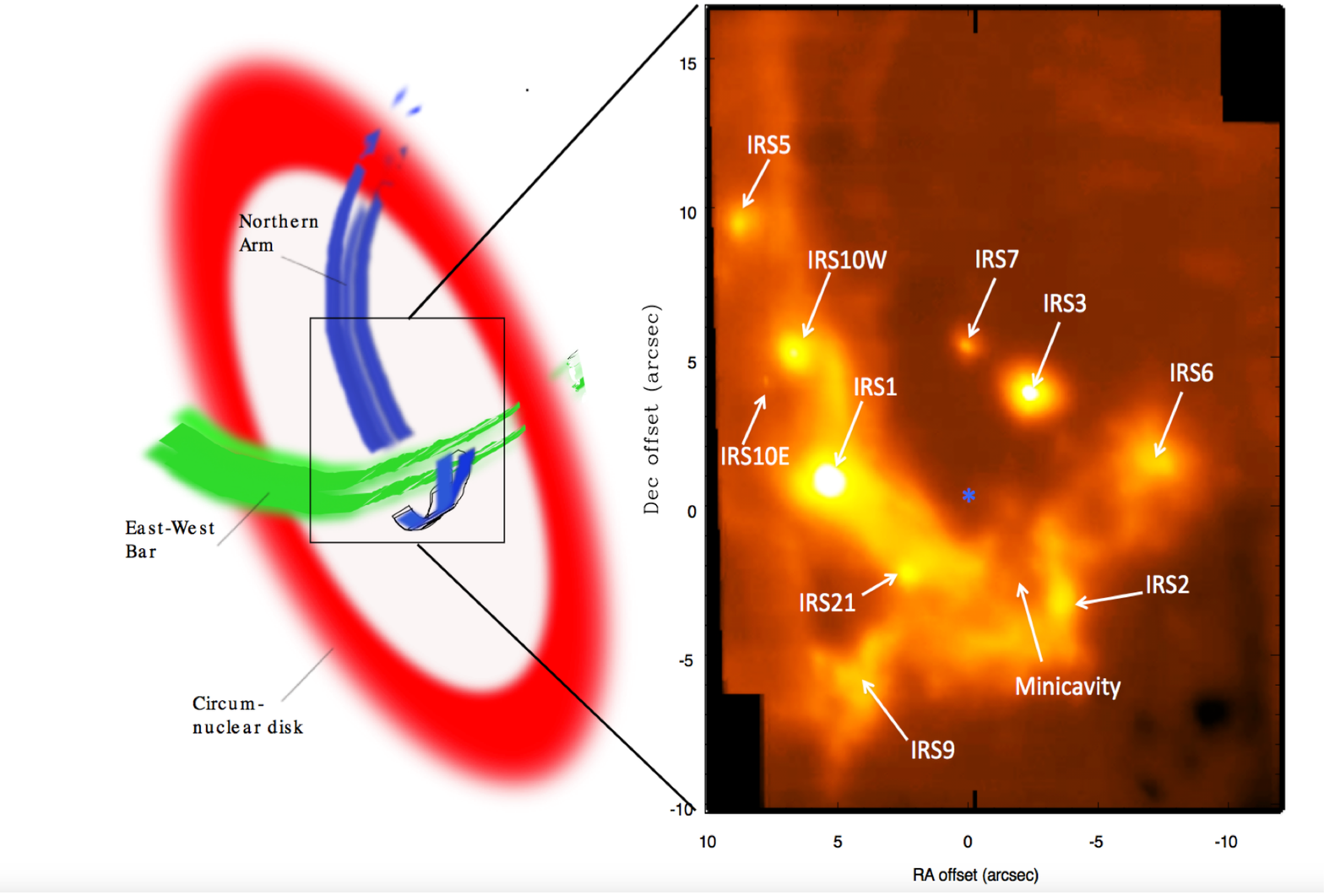}
	\caption{Right: 12.5~$\mu$m acquisition image mosiac.  The positions of several of the bright compact emission peaks discussed in the paper are indicated. Offsets are relative to the position of SgrA*, which is indicated by a star.  Left:  cartoon showing the larger scale structures in the central 5~pc and their relation to the features discussed in the text.} 
	\label{fig:SgrAMos}

\end{figure*}

A mosaic constructed from the acquisition images taken with CanariCam through the 12.5~$\mu$m filter is shown in Fig.~\ref{fig:SgrAMos}.  This image covers a slightly larger area than that used for the polarimetric observations, and serves as a guide to the positions of the compact emission peaks and other structures discussed below.  It also demonstrates the effects of the finite chop used in the polarization measurements.  Negative structures in the south-west corner of the image result from chopping onto the Northern Arm, IRS5 and the compact objects further to the east.  However, the polarization measurements do not extend as far to the south-west as this region, since the emission is too faint and any emission structure in the chopped beam is at very low levels.  The very good agreement with previous observations taken with a range of instruments indicates that contamination from residual emission is not significant. Also shown  in Fig.~\ref{fig:SgrAMos} is a cartoon to illustrate the larger scale structures in the central few parsecs of the Galaxy. 

\section{Reduction}

The observations were reduced using custom Python routines described in \citet{Lopez16}.  
The difference for each chopped pair was calculated and the nod frames then differenced and combined to create a single image for each Position Angle (PA) of the half-wave plate (HWP).  During this process, all images were examined for high or variable background that could indicate the presence of clouds or variable precipitable water vapour, but conditions were good and no data were excluded.   

The ordinary (o-) and extraordinary (e-) rays for each HWP PA were extracted for each slot of every data file. The area extracted from each slot is 1.8 arcsec x 20.0 arcsec, slightly undersized compared to the physical slots to avoid vignetting and edge effects.  Observations at consecutive telescope pointings provided overlap of the slots by about half their width, and the common areas are used for registration in construction of the final mosaic. For each slot of every file, the o- and e-rays, produced by the Wollaston prism, were used to calculate the Stokes parameters I, Q and U according to the ratio method. The percentage polarization is then  P$^2$ = (Q$^2$ +U$^2$)/I$^2$ and P.A. = 0.5arctan(U/Q). The images were checked for consistency in the polarimetric observations between slots, files and Observing Blocks (OBs). Aperture photometry and polarimetric measurement of the brightest objects in the slots were performed for every OB, which yielded variations smaller than 5\% in the Stokes parameters I,Q and U.

CanariCam was mounted on the Nasmyth A platform where the tertiary mirror produces significant instrumental polarization (IP).  The IP was determined during  CanariCam commissioning, and a prescription is provided on the GTC website\footnote[1]{see http://www.gtc.iac.es/instruments/canaricam}.   Specifically, the instrumental polarization is P$_{ins} = 0.6 +/- 0.2 \%$ with a dependence on the PA of polarization given by PA$_{ins}$ = --(RMA+Elev) -- 29.6$^{\rm o}$, where RMA is the Nasmyth rotator mechanical angle and Elev is the telescope elevation; 29.6$^{\rm o}$ is the latitude of the observatory site. 

The instrumental polarization was corrected as follows. The normalized Stokes parameters, q$_{ins}$ = Q$_{ins}$ /I$_{ins}$ and u$_{ins}$ = U$_{ins}$/I$_{ins}$, of the instrumental polarization were computed using the above prescription for the degree P$_{ins}$ and position angle PA$_{ins}$ of the instrumental polarization. Then, q$_{ins}$ and u$_{ins}$ were subtracted from the normalized Stokes parameters of the Galactic Centre images. The polarization efficiency was corrected based on data provided by the GTC website, using the quoted polarization efficiency of 88.3 per cent at 12.5 $\mu$m. For each OB, mosaics were constructed for each slot, taking into account the 1$^{\prime\prime}$ offsets between them. The images through each slot were interpolated to 1/4 pixel,  offsets of 1$^{\prime\prime}$ were applied to the slots and finally slots were co-added. 

The final mosaic of the Galactic Centre was constructed by combining the I, Q and U images for each OB. The images were resized and the mosaics from each OB were registered and shifted to a common pixel, and finally co-added.  The required 0.5 to 2 pixel shifts required are below the typical FWHM of ~0.45$^{\prime\prime}$.  By ensuring that exactly the same offsets are applied to the I, Q, and U images through the slots, the polarization fidelity of the CanariCam instrument is preserved, even if the spatial positioning has some errors.

As a final check on consistency, aperture photometry and polarimetric measurements of prominent features in the images were performed. 
Variations of the total flux counts at a level of 2\% were found, but all images show consistent polarimetric maps, and the final measured polarizations are in agreement with literature values.

\section{Calibration}

\subsection {Position Angle Calibration}
The compact polarized young stellar object AFGL 2591 was used as a PA calibrator.   It is a bright, highly polarized object with P= $4.5 \pm 0.1\%$  at 11.6~$\mu$m, and a wavelength-independent PA across the 8-13~$\mu$m window of $170 \pm 1$ deg (\citealt{Aitken88}). Observations of AFGL 2591 were made at several positions in the slot mask to check the calibration across the CanariCam Field. No significant variations were detected, and  the average P = $4.7 \pm 0.1\%$. at PA= $44.9 ~\pm~ 1.0$ deg. Our measured degree of polarization is in good agreement with those published by  \citet{Aitken88} and \citet{Smith00}.  Comparison of the CanariCam measurements to the literature values indicates  a zero-angle correction of $\Delta$~PA = $125.1 \pm 1.0$ deg, and this correction was applied to all of the observations.

\subsection{Unpolarized Standard}
Brief observations of a bright, and assumed unpolarized star, 37 Sgr (HD175775), were made to provide estimates of the point spread function  and  the instrumental polarization. 37 Sgr was observed in the middle of each slot and the data were reduced following the same procedure employed for the Galactic Centre. After correction for the instrumental polarization outlined above, the average residual percentage polarization was  P$ = 0.24\pm 0.09\%$ indicating that the correction is stable and that the prescription is appropriate.  The observations of AFGL 2591 and 37 Sgr were made on different nights from the observations of the Galactic Centre, and so they can not be used as fiducial PSF references for the Galactic Centre observations, while the flux calibrations are subject to substantial uncertainties. 

\begin{figure*}
\centering
\vspace*{-3.cm} 
	\includegraphics[width=17.5cm]{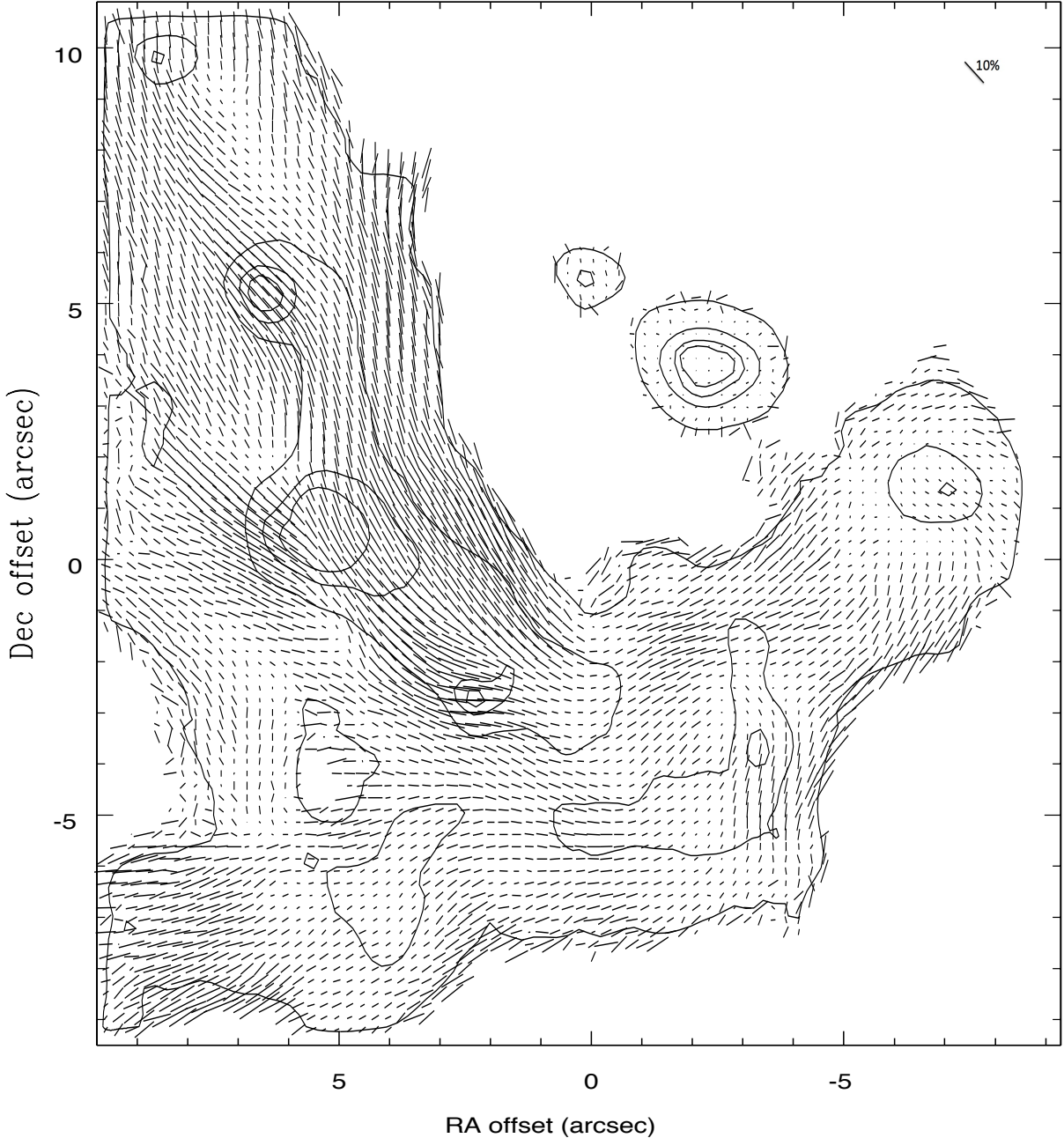}
	\vspace*{-1.5cm} 
	\caption{The emissive polarization map of the central 0.74 x 0.78 pc of the Galaxy.  The length of the polarization vectors is proportional to the percentage polarization, with the scale indicated at the upper right.  The vector position angles have been rotated by 90 deg so that they point along the inferred  magnetic field directions in the plane of the sky.   A flux threshold has been applied to prevent large and spurious vectors from being plotted.  A few contours of intensity are shown to allow registration against the image and the positions of the compact sources identified in fig.\ref{fig:SgrAMos}.   As explained in section 4.2, the flux calibration is uncertain, but the lowest contour is at approximately 1~Jy~arcsec$^{-2}$, with further contours at $\sim$3, 6 and 9~Jy~arcsec$^{-2}$.  Offsets are in arcsec relative to the position of SgrA*.}
		\label{fig:GCPolf2}

\end{figure*}

\subsection{Correction for Interstellar Polarization}

The measured polarization in the Galactic Centre arises from a combination of polarized emission from warm, aligned dust grains in the central parsec together with an interstellar absorptive polarization component produced by dichroic absorption by cool, aligned grains in the interstellar medium between the Earth and the Galactic centre \citep{Aitken86}.  In order to reveal the intrinsic emissive polarization, correction for the absorptive component is required.  The absorptive interstellar polarization component has been estimated by unfolding emissive and absorptive components from spectrophotometric observations across the silicate feature between 8 and 13~$\mu$m, and by measuring the polarization in the 12.8~$\mu$m [Ne II]  emission line, which is assumed intrinsically unpolarised (\citealt{Aitken86}). The isolated compact source IRS3 shows no indication of emissive polarization, with a constant PA across the N-band and a polarization profile compatible with the narrow silicate absorption profile found in lines of sight towards the Galactic centre and other sightlines in the interstellar medium (\citealt{Roche84},  \citealt{Roche85}, \citealt{Bowey98}).  These  analyses indicate that the absorptive component is broadly constant across the central parsec, and is well modelled by the polarization curve derived from the measurements of IRS3 by \citet{Smith00} 

There is evidence of patchy extinction towards sources in the central regions of the Milky Way. In particular, extinction increases markedly in the northern and southern lobes of the circumnuclear disk   (e.g. \citealt{Scoville03}, \citealt{Lau13}), but although the extinction within the central parsec may be enhanced  in some areas, it appears that the interstellar component of the extinction towards the bright emission regions discussed here is not subject to large variations. For example, \citet{Moultaka04} find that the variation over the central 0.5pc is $\Delta A_{K} \le 0.5$mag, while \citet{Schodel10} find that the distribution of extinction derived from H - K band colours is characterised by $A_{Ks} =2.74 \pm 0.3$~mag. \citet{Chatz15} estimated that the extinction by dust within the central few parsecs (e.g. caused by the minispiral) is typically on the order A$_{\rm K} \sim$0.4~mag and probably less than half that in the region studied here.   Similarly, \citet{Roche85} found relatively small variations in the mid-IR absorption optical depth towards the bright compact sources in the central parsec.  It therefore appears safe to conclude that the likely variation in the interstellar absorptive component of polarization at 12.5~$\mu$m in the region considered here is less than 0.1\%. 

The interstellar absorptive polarization is discussed in detail by \citet{Aitken91, Aitken98}, who compare estimates of the interstellar polarization from mid-infrared spectro- and imaging- polarimetry and conclude that all of the results are consistent with a constant polarization of ~1.8\% at PA= $5 \pm 5$ deg at 12.5~$\mu$m. Spectropolarimetry of the isolated dusty supergiant IRS3 \citep{Aitken86} has demonstrated that it has no emissive component of polarization but that it is representative of the interstellar absorptive component  to the Galactic Centre. The measured polarization of IRS3 in the current data is 1.7\% at PA $\sim$10 deg and we have used this value to correct for interstellar polarization. On the assumption that IRS3 is intrinsically unpolarised, this approach corrects for any residual instrumental polarization as well as the interstellar absorptive component.  As pointed out by \citet{Aitken98}, errors of a few degrees in position angle only affect the resultant emissive polarization at the tenth of a percent level.  In making the correction in this way, the polarization on IRS3 is forced to be zero, but it is noteworthy that the residual polarization of the other prominent isolated supergiant IRS7 is less than 1\%, suggesting that this object too has a low intrinsic polarization, and giving confidence that the correction applied is appropriate across the field.

The final polarization image after correction for interstellar polarization, and with the vectors rotated by 90$^{\rm o}$ to indicate the direction of the magnetic field is shown in Figure 2.  The polarization vectors are displayed at intervals of 0.24 arcsec, i.e. with bins of 3 x 3 detector pixels to allow the vector patterns to be seen more clearly and to improve the signal-to-noise ratio per pixel in the lowest flux regions. The resulting vector maps are oversampled by a factor of $\sim$2 with respect to the image resolution. The offsets in the polarization image are with respect to the position of SgrA*, taken as R.A. = 17:45:40.04, dec. = -29:00:28.17 (\citealt{Reid04}).

\begin{figure*}
\centering
	\includegraphics[width=16cm]{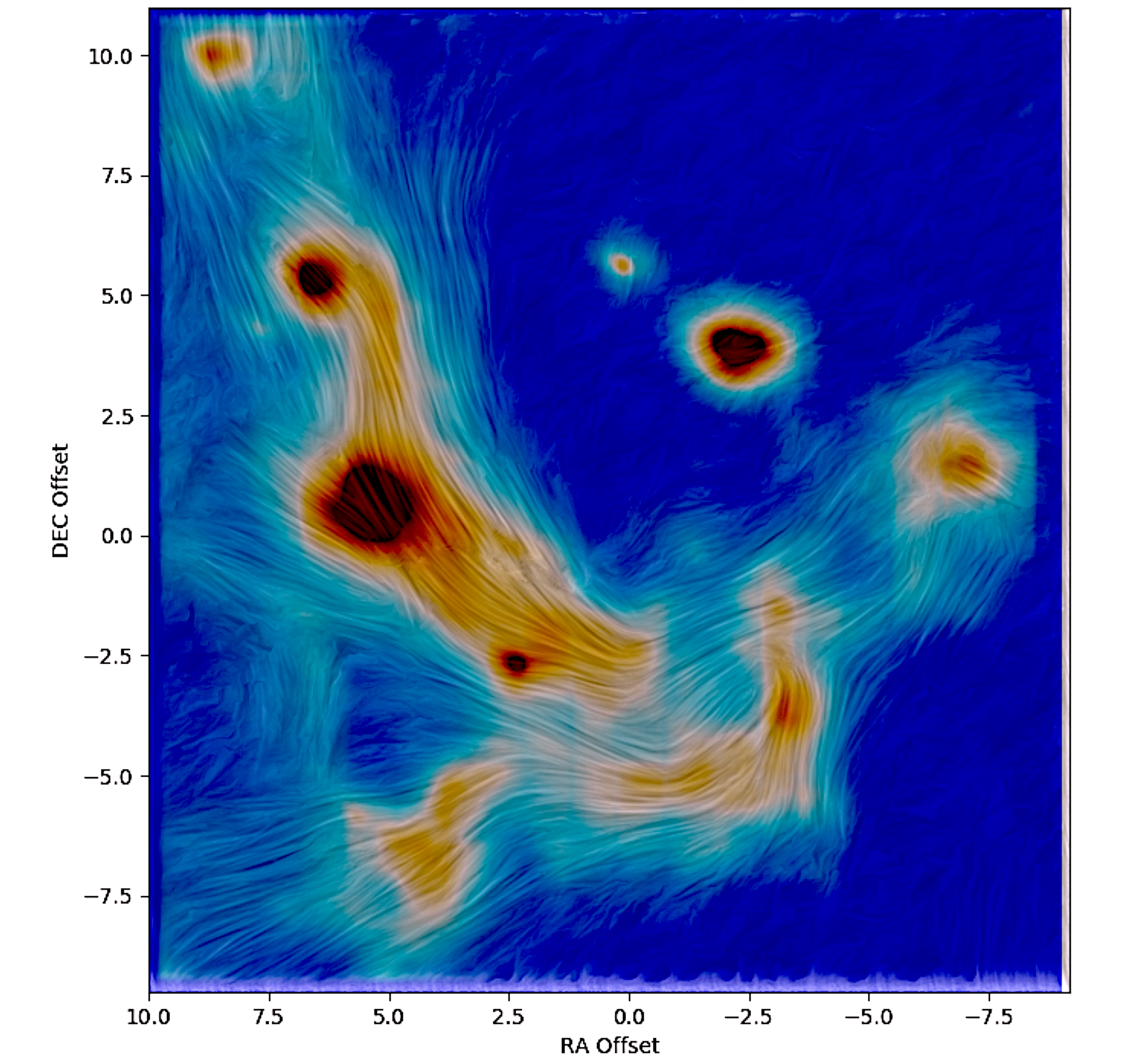}
	\caption{The polarization map of the central 0.75 parsec produced with a line integral convolution to emphasise the coherent structures. }
\label{fig:GCPolf3}

\end{figure*}

\section{Interpretation of Polarization}

Because of dissipative effects in grains spinning in a magnetic field, {\it B}, the grains will tend to align such that the grain spin axis is parallel to its maximum moment of inertia.  The grain spin axis will precess around the magnetic field direction, providing a net alignment of an ensemble of  non-spherical dust grains  such that the grain long axes are preferentially aligned perpendicular to the {\it B} field direction (see e.g. \citealt{Andersson15} for a review). This produces polarized emission with the position angle normal to the {\it B} field, so that the vectors, when rotated by 90$^{\rm o}$, trace out the component of the field directions in the plane of the sky.  The aspect ratios and detailed composition and properties of the emitting grains or the alignment mechanisms in the Galactic Centre are not known, so it is not possible to derive quantitative estimates of the field strength.  However, indirect arguments (see  \citealt{Aitken98}) indicate that the minimum field strength in the bright regions is 2mGauss.

The magnetic field vectors indicated by the polarization measurements presumably reflect the effects of orbital motion and shear coupled with the impacts of outflows from stars in the central regions, together with possible infall.   Previous observations \citep{Aitken91, Aitken98, Glasse03}, have shown that the 12.5~$\mu$m polarization vectors trace out coherent field structures in the Northern Arm with continuous and slowly varying degrees of polarization, suggesting that the field is oriented primarily in the plane of the sky and may be amplified by shearing motions as the material orbits the centre of the Galaxy.  On the other hand, other researchers (e.g. \citealt{Irons12}) have argued that the gas does not move along the streamers, but rather that the prominent Northern Arm structure is produced by orbital precession  of gas in circular orbits  producing a density wave, together with the enhanced ionizing flux from the sources in the central region.  It may be that some of the bright filamentary structures are enhanced by compression by outflowing gas and that they trace ionization fronts at the interfaces the warm material and hot gas. 
  
We note that some alignment mechanisms, such as radiative torques acting on grains might be expected to lead to greater alignment and higher polarization in regions with high radiation fluxes (e.g. \citealt{Andersson15}).  This does not appear to be the case in the central region of the Milky Way, but  \citet{Aitken98} proposed that the grain alignment may be saturated and that variations in the degree of polarization along the Northern Arm reflect changes in the orientation of the field with respect to the line of sight. 

The current results are in agreement with the previously-published mid-IR polarization data, and especially with the most recent results which were presented by \citet{Glasse03}; there is excellent agreement between the two datasets but the higher resolution of the current observations reveal new details in the polarization structure.

\begin{figure*}
\centering
	\includegraphics[width=18cm]{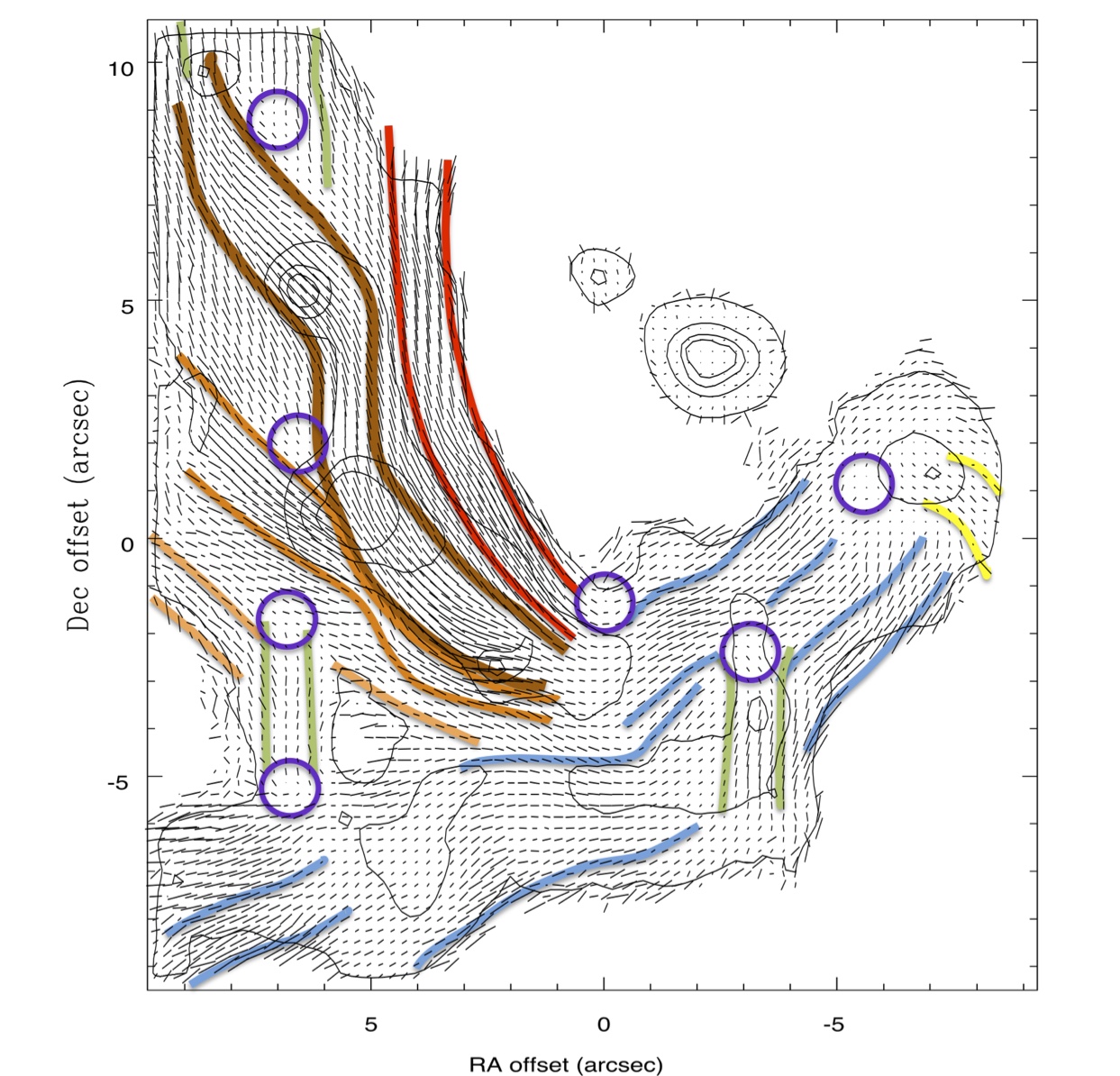}
	\caption{The Polarization map of the central 0.75 parsec from figure \ref{fig:GCPolf2} overlaid with lines to indicate regions with coherent field structures.  Red and brown lines indicate regions of coherent vector orientations on the inner edge of the Northern Arm and in the regions of the compact objects IRS5, IRS10W, IRS1 and IRS21.  Green lines indicate North-South structures, possibly related to the bulk Northern Arm field, while Blue lines trace coherent structures along the East-West bar.    Purple circles indicate null points in the polarization, and generally occur where two coherent polarization structures at different orientations are superposed
  }
\label{fig:GCPolf4}

\end{figure*}

\section{Polarization Structures}

The warm dust emission observed in the mid-infrared lies within the circumnuclear disk traced by molecular lines and far-infrared dust emission.  The emissive polarization from dust in   the CND has been measured at far-infrared and submillimeter wavelengths \citep{Hildebrand93, Novak00} at resolutions of $\sim$40 and 20 arcsec respectively, and indicates that the fields in the CND generally lie in the north-south direction, possibly resulting from wind-up of a poloidal field by the CND rotation.  

The inferred magnetic field in the region of the Northern Arm north of IRS5 is also primarily north-south and is well ordered, consistent with the field lying in the direction of the orbital motion (\citealt{Aitken91,Aitken98,Glasse03}.  However inspection of the data presented in Fig. \ref{fig:GCPolf2} shows that the structure is more complex in the central 0.5 parsec. 
The complexity of the structures and the regions with coherent vectors can be more readily appreciated in Fig.  \ref{fig:GCPolf3}. This figure has been produced with a line integral convolution algorithm \citep{cabral93} which highlights the flow of the field and its filamentary structure in the minispiral, but apparently converging south of SgrA*. 

In Fig \ref{fig:GCPolf4} we have drawn lines to indicate regions of coherent vectors, while circles are drawn at the positions of low polarization or null points.  The lines are drawn by hand and indicate regions that appear to have common or slowly changing polarization structures. In the Northern Arm, regions of north-south vectors are found at the top of the figure and also just above the East-West bar; these are indicated in green and may represent the overall field in the Northern Arm.  Other regions in the Northern Arm show coherent structures that are predominantly north-south on the western rim, but rotating towards 45 degrees as the vectors are traced to the south.  South and east of IRS1, vectors trace out structures with field directions close to 45 degrees.  A run of  vectors connects the bright peaks IRS5, IRS10W, IRS1 and IRS21, suggesting that these sources are linked by a coherent field. 
In contrast, many regions in the East -West bar show coherent structures with PA $\sim 120$ to $140$ degrees, following the overall direction of the E-W bar.   As found previously, the strong ordered vectors in the Northern Arm become less defined and appear to curve below and perhaps wrap around SgrA*, while  the degree of polarization decreases markedly in regions to the west of SgrA*.  

The improved resolution of the data presented here allows us to identify regions of low polarization.  These regions are indicated by circles in figure \ref{fig:GCPolf4} and they generally occur at positions where two coherent structures with different position angles intersect.  For example, the circle to the south-west of IRS5 indicates a null point that occurs where the general N-S field of the Northern Arm intersects with the field that links the bright compact IRS sources along the intensity ridge.  Similarly, a low or null point occurs to the northeast of IRS1, where a stream of vectors at PA $\sim 45^{\rm o}$ intersects with a north-south structure along the ridge linking IRS1 and IRS10W.    It appears that a number of polarized filaments oriented at about 45 deg are superimposed upon the overall N-S run of the magnetic field in the diffuse emission in the Northern Arm.  The most prominent of these features are indicated by the regions that lie between the thick coloured lines drawn in Fig. \ref{fig:GCPolf4}.  A similar situation may prevail in the East-West bar where regions with field directions oriented approximately N-S (e.g. south of IRS2) stand out against the dominant field directions with PA $\sim130^{\rm o}$ in the diffuse material along the bar.

\begin{figure*}
\centering
\vspace*{-3cm} 
	\includegraphics[width=17cm]{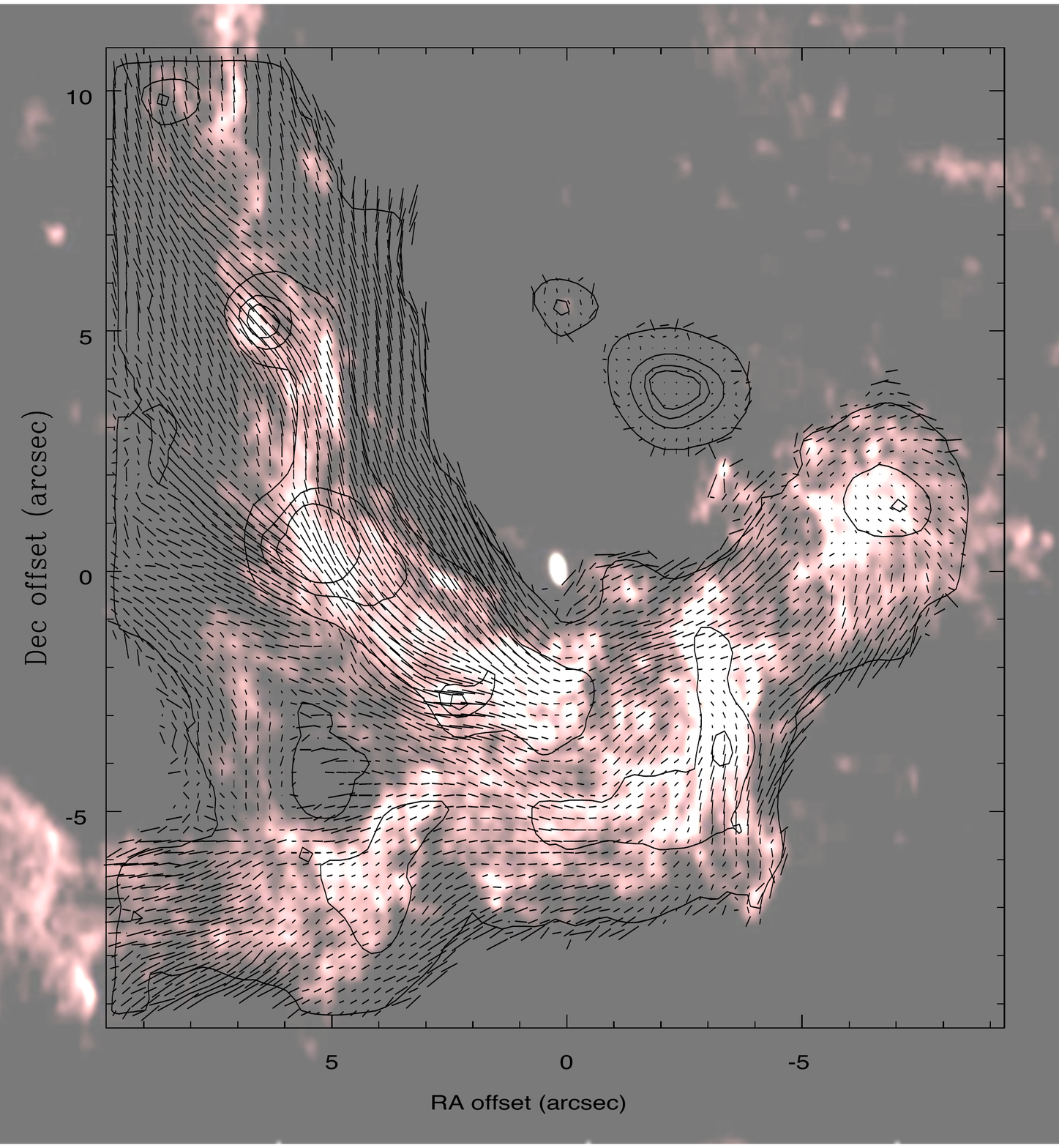}
	\vspace*{-2cm} 
	\caption{The polarization map  superimposed on the 1.3~cm radio map of \citet{Zhao09} displayed as a greyscale, and reproduced with permission of the AAS. The radio image has been aligned manually to the mid-IR intensity images;  the image registration will not be precise, with estimated uncertainties of 0.1 arcsec.  
}
\label{fig:GCPolf5}

\end{figure*}

\subsection{The Compact Sources}
The polarization of the compact IRS sources measured in 6 pixel (0.48 arcsec) diameter apertures is listed in Table \ref{tab:Table1}.
The maximum levels of polarization are found in the region of the Northern Arm to the east of SgrA* (listed as NA Rim in the table) and towards the ridge between IRS1 and IRS21, where the polarization reaches 12\%.  The polarization of the compact sources is similar to, but generally at somewhat higher levels and rotated in angle by a few degrees compared to previous measurements \citep{Aitken86}, reflecting the fact that the small apertures used here integrate over smaller areas, better isolating the emission from the compact objects, and reducing position angle variations within the aperture. The polarization angles listed in Table \ref{tab:Table1} are the measured PAs of the intrinsic emission; i.e. they have not been rotated to indicate the field directions.   The flux densities listed are simply the integrated flux within the aperture, with no correction for underlying diffuse emission.  Comparison with previous estimates of fluxes are difficult because of the small aperture sizes used here and because most published measurements are corrected for interstellar extinction and diffuse background emission (e.g. \citet{Viehmann06}).  Furthermore, the flux calibration was applied from observations of 37 Sgr made on the night before the Galactic Centre polarization  observations, so that while the relative fluxes will be more accurate, the absolute uncertainties are probably at least 20\%, and the flux estimates assume that the psf is similar to that of 37 Sgr. 

In general these measurements emphasise the lack of perturbation of the field by the compact sources. However, there does appear to be dilution of the polarization at the position of IRS10E. IRS10E is visible in Fig.~\ref{fig:SgrAMos} and more prominently in Fig.  \ref{fig:GCPolf3} as a relatively weak compact object approximately 1.5 arcsec south-east of the prominent northern arm object IRS10W.   At this position, there is a reduction in percentage polarization by a factor of $\sim$2 relative to the surrounding areas, but the position angle appears unaffected, consistent with dilution resulting from inclusion of unpolarized flux in the beam.  This is consistent with IRS10E being an unrelated object, separated from the Northern Arm along the line of sight; it is the closest OH-IR star to SgrA* (\citealt{lindqvist}) and has a compact morphology unlike those of the bow-shock sources (\citealt{Sanchez14}).

\begin{table}

	\centering
	Aperture Polarization Measurements 	
	\label{tab:Table1}
	\begin{tabular}{lrrcrr} 
		\hline
		Position & Offset & (arcsec)  &Flux density& P & $\theta$ \\
				&	RA & Dec & Jy&  \% & deg \\
		\hline
		IRS1 & 5.2  & 0.6 &  9.47 & 8.1 & 113.6 \\
		IRS2 & -3.4 & -3.6  & 2.59 & 2.5 & 85.1\\
		IRS5 &  8.6 & 9.9  & 2.43 & 4.8 & 94.1 \\
		IRS7 &0.0 & 5.6  & 1.44 & 0.9 & 99.3 \\
		IRS10W & 6.5 & 5.3 & 4.76 & 9.6 & 130.2 \\
		IRS10E & 7.7 & 4.2 & 0.86 & 3.9 & 108.5 \\
		IRS21 & 2.3 & -2.8 & 3.72 & 10.2 & 87.3 \\
		NA Rim & 2.3  & 0.2 & 0.87 & 12.1 & 121.1 \\
		\hline
	\end{tabular}
	\caption{Emissive polarization and position angle measured in 0.48 arcsec diameter apertures.  Offsets are with respect to SgrA*. }

\end{table}

\subsection{The Northern Arm}

The polarization pattern revealed in these data is complex.  It appears that a number of magnetic filaments are superimposed on the larger scale N-S field. It is striking that the fields do not appear to be randomly distributed, but rather have orientations that are clustered around  distinct position angles.  For example, in the parts of the Northern Arm east of SgrA* and north of IRS21, the field directions  are essentially confined to orientations between 0 and 60 degrees with subregions  on scales of 1-2 arcsec with much more restricted ranges in position angle.   

It is especially noteworthy that, in agreement with the earlier results of \citet{Aitken91, Aitken98}, and with the exception of IRS10E discussed above,  the degree of polarization is unaffected by the intensity peaks, with no evidence of dilution or major disruption of the local field.  This is  demonstrated at higher resolution in the present data. The objects associated with the mid-IR intensity peaks are well studied and appear to be embedded stars with large mass-loss rates (\citealt{Sanchez14,  Tanner05}). The fact that the warm dust associated with the stars is aligned with the same orientation as the diffuse emission in the ridge, and that there is no dilution of polarization by unpolarized or differently oriented polarized circumstellar emission from the stars indicates that they are embedded within the magnetic field lines.  In general, the grain alignment to the magnetic field is a property of the diffuse emission rather than the intensity peaks. The intensity peaks do not appear to have any significant impact on the degree of polarization, although the fact that the bright peaks are linked by coherent field structures suggests that the embedded stars may distort the run of the magnetic field flux tubes on scales of $\sim$0.1 pc.  It seems that the outflows from the stars do not overcome the magnetic field, but rather may compress and shift the field lines.  

Figures 3 and 4 suggest several structures reminiscent of flux tubes.  
The coherent vectors with uniformly high degrees of polarization linking the bright IRS sources from IRS5 at the top of the image through IRS10W, IRS1 and IRS21 before merging into the East-West bar  define the most prominent structure; this is indicated by the brown thick lines in Fig. \ref{fig:GCPolf4}.  High resolution images of these bright compact objects (e.g. \citealt{Sanchez14}, \citealt{Tanner05}, \citealt{Buchholz13}) reveal bow-shock structures that indicate that the outflows from these embedded stars are interacting directly with material in the Northern Arm.  The exact nature of these stars is not clear because their spectra reveal few diagnostics, but it appears that they may be Wolf-Rayet stars with high mass loss rates (\citealt{Tanner05}). We can estimate the impact of those outflows on the filaments:  the energy in the magnetic field is  ${\it B}^2 / 2 \mu \sim 2~10^{-7}$ erg cm$^{-3}$, but the ram pressure from an outflow from a star at a distance of 0.01 pc (0.25 arcsec) with a mass loss rate of $10^{-5}~\rm{M}_\odot$ yr$^{-1}$ and an outflow velocity of 200km s$^{-1}$ is more than an order of magnitude greater.  As discussed above, the absence of perturbation in the polarization pattern in the local field indicates that the stars must be embedded in the polarized material and so the separation between the stars and the diffuse polarized material must be small. The estimates of ram pressure and magnetic energy  suggest that the stellar outflows can compress and push the field lines, consistent with the patterns seen in the polarization vectors in Fig.\ref{fig:GCPolf2} and in the high resolution near-infrared polarization images of \citet{Buchholz13}.
 
Two coherent structures,  indicated by the thick light brown lines in figure \ref{fig:GCPolf4}, cross  the southern part of the Northern Arm with orientation and polarization vectors at PA $\sim 45^{\rm o}$.  One of these appears to flow from the north east to intersect with the structure linking the bright IRS sources just to the east of IRS1 where there appears to be a convergence of flux tubes, and then continues at a similar position angle to the south west. To the east of IRS1, the coherent field appears to join onto a short bright emission filament at the same PA in the radio continuum image in Fig. \ref{fig:GCPolf5} and then either runs parallel to or merges with the ridge of emission linking the IRS sources. A second less defined structure is apparent, approximately  4 arcsec south of this but this becomes indistinct in Fig. \ref{fig:GCPolf4} after it meets a structure with N-S vectors above the East-West bar and is more clearly seen in Fig. \ref{fig:GCPolf3}.  These structures could represent filaments lying, and perhaps tidally stretched,  along orbits around SgrA*, but inclined to the predominantly N-S directions of the brightest parts of Northern Arm (see e.g. the discussions by \citealt{Irons12} and \citealt{Paumard04}, and figure 8 in the latter).  If the magnetic fields do trace the orbits of streamers, sensitive measurements  of a more extended area might permit the estimation of orbital parameters. 

The region of highest emissive polarization is the part of the Northern Arm directly east of SgrA*.  Here the polarization reaches a maximum of 12\%, which is the highest value of dust polarization detected in the mid-IR to date. As discussed by \citet{Aitken98}, if the alignment is saturated. the variations of polarization may reflect the orientation of the field with respect to the line of sight.  The measurements presented here suggest that the maximum polarization is somewhat higher than discussed by \citet{Aitken98}, but they confirm the conclusion that the region of the Northern Arm south and west of IRS1 may lie close to the plane of the sky.

In Figure \ref{fig:GCPolf5}, the polarization vectors are overlaid on the 1.3~cm radio emission map of \citet{Zhao09}.  There is a general but not detailed correspondence between the radio structures and the mid-IR emission, but the higher resolution of the former reveals additional structure. In particular, \citet{Zhao09} in their figure 22 have drawn attention to an apparently helical structure seen most prominently in the region west of IRS5, which they propose may indicate instabilities in a hydromagnetic flow.  This region north of IRS10W coincides with an area of low 12.5~$\mu$m polarization, and with the northernmost null region in Fig.  \ref{fig:GCPolf4}. The null region could mark a turning point in the helical structure, where the projected magnetic field has a sharp change of direction.   The polarization of the loop in the structure immediately north of IRS10W is significantly lower than that in the loop further to the north and above the null region. If the field directions follow the turns in the loop, we would expect a decrease in the net polarization in the region immediately south-west of the null region where the angle of $\sim45^{\rm o}$ is not aligned to the general north-south field of the Northern Arm. The polarization does indeed decrease there, but also stays low as the loop orientation changes to north-south.  There are several possible explanations for reduced polarization along the loop; for example it could result from dilution by emission from a region where the field orientation changes from the plane of the sky to along the line of sight.  High resolution polarization  observations of the region north of IRS5 would enable us to follow the helical structure in more detail and unravel the polarization properties.

\subsection{The East-West Bar}

In figure \ref{fig:GCPolf5}, it is apparent that the polarization in the lower intensity regions of the East-West bar have vectors pointing approximately along the bar and are in general more strongly polarized than the regions with higher radio intensity.   This is especially obvious in the regions to the east and to the south west of IRS9, the region between SgrA* and IRS2, including part of the minicavity, and the region between IRS2 and IRS6.  These areas of high polarization and with vectors aligned along the bar are delineated by the blue lines in Fig. \ref{fig:GCPolf4}.  The behaviour of the polarization in the East-West bar is quite different from that in the Northern Arm; in the latter, the bright sources do not appear to affect the degree or PA of polarization.  The lower levels of polarization in the bright areas in the East-West bar could result from dilution by emission from unpolarized sources or by sources with different polarization position angles.  

South of IRS2, the brightest region in the East-West bar, the polarization vectors are predominately north-south (indicated by the green lines in Fig. \ref{fig:GCPolf4}.  This region, where the gas shows strongly blue-shifted emission (\citealt{Irons12}, \citealt{Zhao09}), is often viewed as a continuation of the Northern Arm.  Models suggest that the material comprising the main structure of the Northern Arm lies behind  SgrA*, but that it loops to the south and west and re-emerges on the near side near IRS2 (\citealt{Irons12}, \citealt{Zhao09}).  The very different orientation of the polarization vectors in this region is consistent with the region around IRS2 being quite distinct from the general structure of the East-West bar, and the ordered field lines suggest that a substantial component of the field lies in the plane of the sky. 

Regions of low polarization in the region around IRS6 in the western part of the bar may result from the superposition of polarized emission with orthogonal directions to the general vector directions where the western arc crosses the bar.    A small area with polarization vectors at PA$\sim 45^{\rm o}$ is indicated in yellow in Fig. \ref{fig:GCPolf4}, which may arise in material associated with the western arc.

\section{Conclusions}

The 12.5~$\mu$m polarization image presented here reveals details of the magnetic field in the central 0.75 parsec of the Galactic centre.  

\noindent
1. In agreement with previous observations, the mid-IR polarization is found to be a property of the diffuse emission rather than the compact, bright sources. 

\noindent
2.  In the Northern Arm, a coherent field structure links the bright embedded stars IRS5, IRS10W, IRS1 and IRS21 and continues south and west of SgrA*.  The embedded objects do not significantly perturb the inferred magnetic field directions in the immediate vicinity and do not dilute the measured degree of polarization, suggesting that the stars are embedded within the field. However, the fact that the most prominent and brightest objects are linked by a coherent run of vectors suggests that the field has been compressed and aligned to the sources  by the stellar outflows.

\noindent 
3.  The maximum polarization of 12\% at 12.5~$\mu$m is found in the Northern Arm east of SgrA*. This may indicate that the magnetic field lies close to the plane of the sky in this region and may represent the maximum polarization produced by the grains. 

\noindent
4  Other coherent field structures are oriented at  PA$\sim 45^{\rm o}$ so that they cross the general run of the Northern Arm. One of these appears to converge with the polarized ridge structure near IRS1, following a filament visible at radio frequencies. These filaments may represent materials on orbits with different inclinations to the bulk of the Northern Arm.

\noindent
5.  In contrast to the Northern Arm, the polarization in the high intensity regions of the East-West bar is generally lower than in the lower intensity regions, while the position angles also deviate from those in the surrounding diffuse material.  It appears that the low intensity regions trace the predominant magnetic field directions, which run along the bar (with PA $\sim 130^{\rm o}$), but that the polarization is diluted or disrupted in the bright regions. It seems that the bright structures in the East-West bar are not embedded in the larger-scale east-west magnetic field in the same way as the sources in the Northern Arm are embedded in their magnetic filament.  In the western part of the bar, the polarization may be diluted by the superposition of fields from material associated with the western arc at PA $\sim 45^{\rm o}$, while the emergence of the southern extension of the Northern Arm may account for the region of north-south vectors  south of IRS2. 

\noindent
6. A helical structure identified by \citet{Zhao09} and proposed to arise through instabilities in a magnetic medium appears to have some effect on the polarization vectors. For example, the projected turning point of a loop in the helix coincides with a null in polarization.    More coverage of this structure to the north of the present image would allow a more detailed comparison and an investigation of the field directions. 

\section*{Acknowledgments}

 This paper is based on observations obtained with the Gran Telescopio Canarias (GTC), installed in the Spanish Observatorio del Roque de los Muchachos of the Instituto de Astrofisica Canarias, in the island of La Palma. We are very grateful to the support staff, and particularly to the dedicated help of Carlos Alvarez who provided excellent support and ensured that good quality data were obtained.  We are also grateful to the whole CanariCam team for their skill in delivering a cryogenically cooled dual beam mid-ir imager-spectrometer-polarimeter.  C.M.T. acknowledges support for this research from NSF awards AST-0903672, AST-0908624, and AST-1515331.

{}

\label{lastpage}


\begin{thebibliography}{}

\bibitem[\protect\citeauthoryear{Aitken et al. }{1986}]{Aitken86}  Aitken D.K., Roche P.F., Bailey J.A., Briggs G.P., Hough J.H., Thomas J.A.,1986, MNRAS, 218, 363
\bibitem[\protect\citeauthoryear{Aitken et al. }{1988}]{Aitken88}  Aitken D.K., Roche P.F., Smith C.H., James S.D., Hough J.H., 1988, MNRAS, 230, 629
\bibitem[\protect\citeauthoryear{Aitken et al. }{1991}]{Aitken91}Aitken D. K., Gezari D., Smith C. H., McCaughrean M., Roche P. F., 1991, ApJ, 380, 419
\bibitem[\protect\citeauthoryear{Aitken et al. }{1998}]{Aitken98}  Aitken D.K., Smith C.H., Moore T.J.T., Roche P.F., 1998, MNRAS, 299, 743
\bibitem[\protect\citeauthoryear{Andersson, Lazarian  \& Vaillancourt }{2015}]{Andersson15} Andersson B-G, Alazarian A., Vaillancourt J.E., 2015, ARAA 55, 501
\bibitem[\protect\citeauthoryear{Boehle et al.}{2016}]{Boehle16} Boehle A., et al. 2016. ApJ, 830, 17  
\bibitem[\protect\citeauthoryear{Bowey, Adamson \& Whittet}{1998}]{Bowey98} Bowey, J. E., Adamson, A. J., Whittet, D. C. B., 1998. MNRAS 298, 131
\bibitem[\protect\citeauthoryear{Buchholz et al.}{2013}]{Buchholz13} Buchholz R.M, Witzel G., Sch{\"o}del R., Eckart A., 2013, A\&A, 557, 82
\bibitem[\protect\citeauthoryear{Cabral \& Leedom}{1993}]{cabral93}  Cabral B., Leedom L., 1993. Proc SIGGRAPH '93 263
\bibitem[\protect\citeauthoryear{Chatzopoulos et al.}{2015}]{Chatz15}  Chatzopoulos S., Gerhard O., Fritz T.K., Wegg C., Gillessen S., Pfuhl O., Eisenhauer F., 2015, MNRAS, 453, 939
\bibitem[\protect\citeauthoryear{Genzel, Eisenhauer \& Gillessen}{2010}]{Genzel10} Genzel R., Eisenhauer F.,  Gillessen S., 2010. RevModPhys 82, 312
\bibitem[\protect\citeauthoryear{Gezari \& Yusef-Zadeh}{1991}]{Gezari91} Gezari D., Yusef-Zadeh F., 1991, 'Astrophysics with Infrared Arrays', ed R Eston, ASP Conf Ser14, 214
\bibitem[\protect\citeauthoryear{Glasse, Aitken \& Roche}{2003}]{Glasse03} Glasse A.C.H., Aitken D.K., Roche P.F., Astron Nachr 324, 3
\bibitem[\protect\citeauthoryear{Hildebrand et al.}{1993}]{Hildebrand93} Hildebrand R.H., Davidson J.,A., Dotson J., Figer D.F., Novak G., Platt S.R., Tao L., 1993. ApJ, 417, 565
\bibitem[\protect\citeauthoryear{Irons, Lacy \& Richter}{2012}]{Irons12} Irons W.T., Lacy J.H., Richter M.J., 2012, ApJ 755, 90
\bibitem[\protect\citeauthoryear{Jackson et al}{1993}]{Jackson93} Jackson J.M., Geis N., Genzel R., Harris A.I., Madden S., Poglitsch A., Stacey G.J., Townes C.H., 1993, ApJ, 402, 173
\bibitem[\protect\citeauthoryear{Lacy, Achtermann \& Serabyn}{1991}]{Lacy91} Lacy, J. H., Achtermann, J. M.,  Serabyn, E. 1991, ApJ, 380, 71
\bibitem[\protect\citeauthoryear{Lau et al.}{2013}]{Lau13} Lau R.M., Herter T.L., Morris M.R., Becklin E.E., Adams J.D., 2013, ApJ, 775, 37
\bibitem[\protect\citeauthoryear{Lindqvist et al.}{ 1992}]{lindqvist}Lindqvist M., Winnberg A., Habing H.J.,  Matthews H.E. 1992, A\&AS, 
92, 43
\bibitem[\protect\citeauthoryear{Lopez-Rodriguez et al.}{2016}]{Lopez16} Lopez-Rodriguez E., Packham C., Roche P.F., Alonso-Herrero A., Diaz-Santos T. et al. 2016, MNRAS 458, 3851
\bibitem[\protect\citeauthoryear{Moser et al.}{2017}]{Moser17} Moser L., et al., 2017,  A\&A 603, 68
\bibitem[\protect\citeauthoryear{Moultaka et al.}{2004}]{Moultaka04} Moultaka J., Eckart A.,  Viehmann T., Mouawad N., Straubmeier C., Ott T., Sch{\"o}del R., 2004, A\&A, 425, 529
\bibitem[\protect\citeauthoryear{Novak et al.}{2000}]{Novak00} Novak G., Dotson J.L., Dowell C.R., Hildebrand R.H., Renbarger T., Schleuning D.A., 2000, ApJ, 529, 241
\bibitem[\protect\citeauthoryear{Packham, Hough \& Telesco}{2005}]{Packham2005} Packham, C., Hough J.H., Telesco C., 2005. 'Astronomical Polarimetry', ASP Conf Ser 343, 38
\bibitem[\protect\citeauthoryear{Paumard, Maillard \& Morris}{2004}]{Paumard04} Paumard, T., Maillard, J-P., Morris, M., 2004, A\&A, 426, 81
\bibitem[\protect\citeauthoryear{Reid \& Brunthaller}{2004}]{Reid04} Reid M.J., Brunthaler A., 2004, ApJ, 616, 872
\bibitem[\protect\citeauthoryear{Roberts \& Goss}{1993}]{Roberts93} Roberts, D. A., Goss, W. M. 1993, ApJS, 86, 133
\bibitem[\protect\citeauthoryear{Roche \& Aitken}{1984}]{Roche84}  Roche P.F., Aitken D.K., 1984, MNRAS, 208, 481
\bibitem[\protect\citeauthoryear{Roche \& Aitken}{1985}]{Roche85}  Roche P.F., Aitken D.K., 1985, MNRAS, 215, 425
\bibitem[\protect\citeauthoryear{Sanchez-Bermudez et al.}{2014}]{Sanchez14} Sanchez-Bermudez  J., Sch{\"o}del  R., Alberdi A., Muzic K.,  Hummel C.A., Pott J.-U. 2014, A\&A  567, 21
\bibitem[\protect\citeauthoryear{Scoville et al}{2003}]{Scoville03} Scoville N.Z., Stolovy S.R., Rieke M., Christopher M.H., Yusuf-Zadeh, F., 2003. ApJ,594, 294
\bibitem[\protect\citeauthoryear{Sch{\"o}del et al}{2010}]{Schodel10} Sch{\"o}del R., Najarro F.,  Muzic K., Eckart A.,   2010, A\&A, 511, A18
\bibitem[\protect\citeauthoryear{Serabyn \& Lacy}{1985}]{Serabyn85} Serabyn, E., Lacy, J. H.,  1985, ApJ, 293, 445
\bibitem[\protect\citeauthoryear{Serabyn et al}{1988}]{Serabyn88} Serabyn, E., Lacy, J. H., Townes, C. H., Bharat, R. 1988, ApJ, 326, 17
\bibitem[\protect\citeauthoryear{Smith, Aitken \& Roche}{1990}]{Smith90}  Smith C.H., Aitken D.K., Roche P.F., 1990, MNRAS, 246, 1
\bibitem[\protect\citeauthoryear{Smith et al}{2000}]{Smith00}  Smith C.H.,  Wright C.M., Aitken D.K., Roche P.F., Hough J.H., 2000, MNRAS, 312, 327
\bibitem[\protect\citeauthoryear{Tanner et al.}{2005}]{Tanner05} Tanner, A., Ghez, A., Morris, M.R., Christou, J.C., 2005. ApJ, 624, 742
\bibitem[\protect\citeauthoryear{Telesco, Davidson \& Werner}{1996}]{Telesco96} Telesco C. M., Davidson J. A., Werner M. W., 1996, ApJ, 456, 541
\bibitem[\protect\citeauthoryear{Telesco et al.}{2003}]{Telesco03} Telesco C.M., Ciardi D., French J., Ftaclas C.,  Hanna K.T., Hon D.B., Hough J.H., Julian J.A., Julian R., Kidger M., Packham C., Pina R.K., Varosi F., Sellar R.G., 2003, in Iye, M, Moorwood A.F.,  (eds.), Vol. 4841 of Society of Photo-Optical Instrumentation Engineers (SPIE) Conference Series, p913
\bibitem[\protect\citeauthoryear{Viehmann et al}{2006}]{Viehmann06}  Viehmann T., Eckart A., Sch{\"o}del R., Pott J-U., Moultaka J., 2006, ApJ, 642, 861  
\bibitem[\protect\citeauthoryear{Zhao et al}{2009}]{Zhao09}  Zhao, J-H., Morris, M. R., Goss, W. M., An, T. 2009, ApJ, 699, 186
\bibitem[\protect\citeauthoryear{Zhao et al}{2010}]{Zhao10}  Zhao, J-H., Blundell R., Downes D., Schuster K.F., Marrone D.P., 2010, ApJ, 723, 1097


\bsp

\end{thebibliography}
\end{document}